# Post-Heat Treatment Design for High-Strength Low-Alloy Steels Processed by Laser Powder Bed Fusion

Soumya Sridar, Yunhao Zhao, Kun Li, Xin Wang, Wei Xiong*

*Physical Metallurgy and Materials Design Laboratory, Department of Mechanical Engineering and Materials Science, University of Pittsburgh, Pittsburgh, PA 15261, USA.*

## Abstract

A novel post-heat treatment design is implemented for additively manufactured copper-bearing high-strength low-alloy (HSLA) steels by understanding the processing-structure-property relationships. Hot isostatic pressing (HIP) is adopted in order to reduce the porosity from 3% to less than 1% for the HSLA-100 steel processed using laser powder bed fusion (LPBF). Quenching dilatometry is employed to design the parameters for the HIP cycle with an optimized cooling process. In order to achieve the maximum amount of martensite, drop cooling after HIP is found to be more suitable than controlled cooling. A subsequent cyclic re-austenitization is introduced to achieve effective grain refinement to compensate for the coarsened microstructure after HIP. The re-austenitization effectively leads to a 60% reduction in the prior austenite grain (PAG) size. The microstructure of the as-built and HIP HSLA steels before and after cyclic re-austenitization consists of martensite, bainite, and martensite/retained austenite (M/A) islands. Tempering heat treatment is applied after HIP, to induce strengthening due to precipitation hardening, that is optimized through microhardness and microstructure characterization. The peak hardness is achieved at 5 hours of aging and the microstructure consists of tempered martensite, bainite, and M/A islands. A significant increase in the tensile yield strength, as well as the ductility, is achieved in comparison with the as-built alloy with the tailored microstructure obtained after the application of the designed post-heat treatment to HSLA-100 steels processed using LPBF. As the first attempt on LPBF of HSLA steel, this work indicates the significance of post-heat treatment design for the AM technique.

* Corresponding author: *Tel: +1-(412) 383-8092, Tel: +1-(412) 624-4846*
*Email: weixiong@pitt.edu or w-xiong@outlook.com*

**1. Introduction**

Additive manufacturing (AM) is a process that is capable of fabricating structural parts layer by layer using 3D (three dimensional) numerical models [1]. In recent years, AM has received great attention due to its ability to produce complex-shaped objects with high efficiency and less environmental impact [2–4]. One of the commonly used AM techniques is the laser powder bed fusion (LPBF) process which involves melting successive layers of powder locally using a high-intensity beam [5]. Materials processed using LPBF form fine-grained microstructures as a result of the interaction between the high energy laser beam and the metallic powders leading to high heating and cooling rates as well as rapid solidification [6]. It has been reported that AM builds can generally exhibit higher strength in comparison with their corresponding wrought alloys, due to the strengthening achieved by the fine microstructure [7].

Copper containing high-strength low-alloy (HSLA) steels are widely used in naval applications such as submarines and offshore structures, line pipes and hull of ships [8]. It is also of great interest for other structural applications such as automobiles, mining instruments, and railways [9]. These applications require a combination of high strength and toughness as well as good weldability. An increase in the yield strength from 55 ksi (379 MPa) to 80 ksi (552 MPa) led to a weight reduction to nearly 800 tons in shipbuilding [10]. Besides, steels that can be welded with a lesser constraint on welding parameters and without any requirement for preheating is desirable as it leads to a considerable reduction in the processing cost [11]. In general, high carbon steels such as HY80 were employed for naval applications, which requires highly controlled welding parameters and preheating during fabrication leading to poor weldability and high fabrication cost. They were eventually replaced by HSLA-100 (100 denotes the minimum obtainable tensile yield strength in ksi) steels that are cheap and do not require any preheat or stringent parameters during welding [10]. In view of the high strength and toughness, the cost and weight-effective HSLA-100 steels are suitable candidate material for shipbuilding.

In HSLA-100 steels, the carbon content was limited to be < 0.06 wt.%, so that the composition is located in the Zone I of the Graville weldability diagram and hence, the susceptibility to hydrogen-induced cracking is reduced [12]. This leads to superior strength and toughness in the heat-affected zone of the weldments, thus, achieving excellent weldability [13]. In order to balance the loss of strength due to the lower carbon concentration, copper was added to improve the strength with the help of precipitation hardening [14]. To compensate for the hot shortness caused by the segregation of copper in the grain boundaries, nickel was added [15]. In addition to this, molybdenum was added in minor amounts to achieve an increase in toughness with the precipitation of M$_2$C (M: Mo as major species, Cr and Fe are minor) precipitates [16]. Synchrotron XRD studies for HSLA steels have shown that the increase in impact toughness is due to the dissolution of cementite, which provides the necessary carbon for the precipitation of the M$_2$C phase [17]. Niobium was also added in small amounts to induce the precipitation of NbC carbide. These precipitates were found to inhibit the grain growth during austenitization, to achieve high strength as well as toughness [18]. Due to the outstanding mechanical properties and weldability exhibited by copper-containing HSLA steels such as HSLA-100, they are potential candidate material for applying AM. However, there are no reports available for AM of copper-bearing HSLA steels so far.

In this work, an attempt is made to perform LPBF and design an appropriate post-heat treatment to achieve the design targets for copper-bearing HSLA-100 steel. Hot isostatic pressing (HIP) has been adopted to mitigate the porosity, followed by cyclic re-austenitization, in order to achieve superior mechanical properties in the AM HSLA steels. With the aim of understanding the structure-property relationship, extensive microstructure characterization has been performed, in order to correlate the phase evolution with the mechanical properties during various stages of heat treatment design. It should be noted that this is the first attempt of processing HSLA-100 steel using the LPBF method.

It is obvious that the optimization of the printing parameters is equally important as the post-heat treatment design. Recently, several studies have been reported where physics-based analytical models are developed and verified to establish various characteristics of the LPBF process [19–23]. The in-process temperature was estimated by deducing the melt pool evolution and a positive correlation was found between the melt pool and the powder packing density [19]. The 3-dimensional temperature distribution during scanning was determined using the point moving heat source solution [20]. The porosity of the built part was estimated using the dimensions of the melt pool [21]. Using sensitivity analysis, the lack-of fusion porosity was found to increase with increasing hatch spacing and layer thickness [22]. The post-printing average grain size decreases with increase in laser power and scanning speed [23]. From these observations, it can be inferred that analytical models are beneficial in simulating the printing parameters of the LPBF process.

At present, we report the efforts for the post-heat treatment design on the as-built alloys processed using EOS M 290 Machine (EOS company, Germany) using the default printing parameters for stainless steel 316L. Further optimization of the printing parameters for HSLA-100 steels to achieve dense AM builds is the focus of our future studies. As a consequence, only tensile strength is studied in this work, and it is planned that other mechanical properties will be further reported together with the optimized laser melting parameters and post-heat treatment.

**2. Systems Design Chart of AM HSLA steels**

A systems design chart demonstrates how hierarchical structural features contribute to the properties along with the evolution of the structure during different processing conditions [24–26]. Figure 1 shows the systems design chart of HSLA-100 steel processed using AM with the identified process-structure-property linkages. Each connection between process, structure, and property indicates the major correlation between these attributes. In order to achieve optimum mechanical properties and corrosion resistance in HSLA-100 steel, the process and the corresponding microstructure needs to be tailored appropriately. High yield strength and excellent resistance to stress corrosion cracking can be achieved in AM HSLA-100 steels by designing a suitable post-heat treatment process. Defects such as porosity and inclusions need to be reduced as much as possible during this process. Porosity is found to have a notable deleterious effect on the mechanical properties such as tensile strength and elongation [27]. It has been reported that defects such as pores, oxidation, and segregation that are normal to the loading direction act as local stress raisers, which reduce the tensile yield strength [28,29]. In addition to this, pores within the AM build affect the corrosion resistance considerably [30]. These pores are considered as pits that can grow easily leading to pitting and the corrosion rate within the pores can build up severe electrolytic conditions leading to a stable pit growth. Furthermore, the presence of either residual or applied stresses can lead to stress corrosion cracking in the as-built AM components [31]. Hence, efforts to eliminate porosity is considered as one of the important steps for quality control in AM builds.

A feasible solution to reduce porosity in the as-built AM components is to perform HIP with high pressure at high enough temperature in an inert gas atmosphere. The applied pressure leads to the collapse of the pores through plastic flow and material transport [32]. It has been confirmed that gas pores, lack-of-fusion as well as large-scale internal porosity can be removed using HIP in AM components [33]. There are several reports where HIP has been suggested as an important step during post-heat treatment to increase the density of as-built AM parts [33–36]. However, the major shortcoming in the HIP process is the grain coarsening due to the slower cooling rates. Hence, grain refinement after HIP with the help of a post-heat treatment process is inevitable for AM components.

Generally, it is not appropriate to apply any complicated deformation technique on AM builds to refine the prior austenite grains (PAG) for fine microstructure. Techniques such as ausforming have been applied previously to reduce the PAG [37] size, which will lead to change in the shape of the component. Hence, a post-heat treatment that can initiate grain refinement without deformation of the printed component needs to be employed. One of the feasible methods is to perform cyclic re-austenitization [38,39], which involves cycles of short time austenitization above the $Ac_3$ temperature to form complete austenitic structure, followed by quenching. With the application of the short-term heat treatment cycles for several times, the dislocation density will increase during each cooling cycle as the structure transforms from austenite to martensite, and this will significantly increase the nucleation potency of martensite to achieve a refined microstructure [40]. Hence, cyclic re-austenitization is a distortion-free and cost-effective post-heat treatment process to achieve grain refinement after HIP of AM builds.

The resultant microstructure after the post-heat treatment comprising of HIP and cyclic re-austenitization should consist of a fine-grained bainite/martensite matrix as a result of the fine PAG. The presence of Zener pinning particles will refine the PAG size effectively during the cyclic re-austenitization. Further increase in strength can be achieved by applying tempering heat treatment after cyclic re-austenitization that will impart precipitation strengthening to AM HSLA-100 steels. Tempering heat treatment facilitates the precipitation of a high fraction of BCC Cu clusters and $M_2C$ precipitates at the optimum time. As a result of the above-mentioned heat treatment processes applied and microstructures developed, desirable properties such as yield strength greater than 100 ksi (690 MPa) and elongation greater than or equal to 16% are the key design targets to produce AM HSLA-100 steels potentially with comparable or even exceeded properties than cast HSLA-100. Additionally, superior weldability originating due to the presence of a low amount of carbon and other alloying elements makes HSLA-100 steel a suitable candidate material for AM processes. Further information relevant to the design requirements for HSLA-100 steels can be found elsewhere [10,13].

## 3. Materials and Methods

### 3.1. LPBF processing

The composition of the HSLA-100 steel powder used for the LPBF process was modified based on the cast HSLA-100 composition using a high-throughput CALPHAD-based ICME (CALPHAD: Calculation of Phase Diagrams, ICME: Integrated Computational Materials Engineering) modeling technique, which will be described elsewhere [41]. Argon gas-atomized HSLA steel powder with composition (in wt.%) of Al: 0.006, C: 0.046, Cr: 0.4, Cu: 1.44, Mn: 0.9, Mo: 0.8, Nb: 0.03, Ni: 3.47, Si: 0.19 was manufactured by Praxair Co., within a mesh size of -200 (74 μm) to -325 (44 μm). The composition of the manufactured steel powder was close to the optimum composition obtained from the CALPHAD-based ICME composition screening. Cubes (10 x 10 x 10 mm$^3$) and tensile bars (according to ASTM E8 standard) were printed using the EOS M 290 machine (EOS company, Germany) with the factory default parameters for stainless steel (SS) 316L (Power=195 W; Scan speed=1083 mm/s; Hatch spacing=0.9 mm; Layer thickness=0.2 mm, Scan pattern= parallel stripes, Energy density= 100 J/mm$^3$).

### 3.2. Thermodynamic calculations

Thermodynamic calculations were carried out using Thermo-Calc software [42] with the TCFE9 database for the HSLA steel powder composition. The Ac$_3$ temperature is 754°C according to the calculated step diagram (phase fraction vs. temperature plot) shown in Fig. 2. Hence, the HIP temperature was chosen as 950°C, which is nearly 200°C above the calculated Ac$_3$ temperature to ensure complete austenitization. Also, it can be observed that at 550°C, the model-predicted phase fractions of Cu and M$_2$C precipitates were nearly the highest, and hence, it was chosen as the tempering temperature in order to achieve maximum precipitation hardening.

### 3.3. Quenching dilatometry

Quenching dilatometry was used for parameter optimization of the HIP with two different cooling processes after isothermal heat treatment. The first one is controlled cooling, where the sample is cooled with a constant cooling rate of 10 K/min and the other one is drop cooling, where the furnace is switched off after isothermal heating. Based on the optimum homogenization time for additively manufactured HSLA steels [43], the total time for heating from the 754°C (calculated Ac$_3$ temperature) to 950°C, holding at 950°C and cooling from 950°C to 754°C was set. Thus, the holding times were calculated to be 45 and 60 minutes for controlled and drop cooling, respectively. In order to identify the optimum cooling process and quantify the impact of cooling on microstructure evolution after HIP, two different temperature profiles as shown in Fig. 3 are considered in the quenching dilatometry of AM HSLA steel. Cuboids of dimensions 5 x 5 x 10 mm were cut from the printed HSLA steel samples using electric discharge machining (EDM, Mitsubishi MV2400S, Japan). The dilation as a function of temperature was measured using TA DIL805 D dilatometer (TA Instruments, Inc, USA) where the sample was held between two quartz rods connected to the linear voltage differential transducer to determine the dilation change with the accuracy up to 50 nm. An S-type thermocouple was spot welded in the middle of the sample surface for temperature determination. The samples were heated using an induction coil and cooled using Helium gas. A vacuum below $10^{-4}$ bar was maintained while heating to and holding at the austenitization temperature.

### 3.4. HIP and cyclic re-austenitization

After identifying the optimal cooling process using dilatometry, the printed tensile bars were subjected to HIP (AIP18-30H by American Isostatic Presses Inc.) at 950°C with a pressure of 30 ksi (~ 207 MPa) and a holding time of 60 mins followed by drop cooling. Subsequently, cyclic re-austenitization was applied to the HIP HSLA samples to attain microstructure refinement. A single re-austenitization cycle comprises of isothermal holding of the HIP HSLA sample at 950°C for 1 min followed by water quenching. Two cycles of re-austenitization were performed based on our results for the optimum number of cycles required for achieving maximum grain refinement in additively manufactured HSLA steels [44]. The temperature-time cycle for the HIP with the cyclic re-austenitization process is given in Fig. 3(b). The density of the as-built and HIP HSLA samples were measured using the Archimedes method. According to the calculated phase fraction of HSLA-100 shown in Fig. 2, the total phase fraction of $M_2C$ and Cu will reach the maximum at 550°C without forming other detrimental precipitates. Therefore, in order to further identify the peak hardening effects in the tempering process, isothermal heat treatment at 550°C for three different durations (3, 5, and 7 hours) followed by water quenching was employed for HIP HSLA steels before and after cyclic re-austenitization.

### 3.5. Microstructure characterization

In this study, all the microstructure characterization was performed in the build direction (i.e., X-Z plane, X, Y and Z denotes the normal (scan), transverse (print) and build directions, respectively) of the sample, which develops an anisotropic microstructure as well as properties, thus weakening the mechanical performance. The melt pool that forms during the AM process has the highest thermal gradient along the direction perpendicular to its boundary [45], which corresponds to the X-Z plane of the AM build. Therefore, the microstructure and properties in the build direction will be significantly affected due to this gradient, in comparison with the other planes of the build. All the samples were ground from 400 to 1200 grit SiC emery paper and cloth polished with diamond and alumina suspensions containing 1 and 0.01 μm particles, respectively. The etching was done with 2% Nital (2 ml $HNO_3$ + 98 ml ethanol) by immersing the sample surface for 10-30 seconds.

The microstructure of the etched samples was viewed with Zeiss Axio Lab A1 optical microscope and Zeiss Sigma 500 VP SEM (scanning electron microscope) with a field emission gun (FEG) source. Further detailed phase analysis was performed with EBSD (electron backscattered diffraction) using FEI Scios Dual-Beam FIB-SEM, with a step size of 0.08 μm. The EBSD data was analyzed using TSL-OIM software Version 8. The prior austenite grains (PAG) were reconstructed from the EBSD data using the ARPGE software package [46,47]. An algorithm based on the orientation relationship between the austenite and martensite/bainite is used to determine the variants directly inherited by a single prior austenite grain. The orientation relationship (OR) proposed by Greninger and Troiano [48] was used since the highest percentage of reconstruction was obtained using this OR in comparison with the Kurdjomov-Sachs [49] and Nishiyama-Wassermann [50,51] OR. The size of the reconstructed PAG was measured using the linear intercept method [52].

### 3.6. Hardness and tensile tests

Hardness measurements were carried out in the as-polished X-Z plane using the Vickers microhardness tester (LM800, LECO Corporation, USA) with a load of 100 g and a dwell time of 10 sec. The reported values are an average of ten readings. Tensile tests were carried out at room temperature using MTS 880 universal testing machine with 100 kN capacity. A 25 mm gauge length extensometer was used for measuring the displacement and the tests were conducted with a strain rate of 0.03 $min^{-1}$. The applied load was oriented parallel to the Y-direction of the tensile bar.

## 4. Results and Discussion

### *4.1. HSLA-100 steel processed by LPBF*

The relative density of the as-built HSLA steel was determined to be nearly 97% using the Archimedes principle. The higher amount of porosity in the as-built HSLA samples can be attributed to the parameters used, i.e., applying the EOS factory developed printing parameters for SS316L to build HSLA samples. Further efforts on optimization of the *in-situ* melting parameters of HSLA-100 will be reported elsewhere. The optical and SEM micrographs along the build direction of the as-built HSLA sample are shown in Fig. 4. It is clear that the microstructure is fine which is typical for an alloy processed using AM. The optical micrograph shows the presence of dark and light phases. Similarly, the SEM micrograph exhibits two different phase morphologies, one with laths and the other without visible laths. The former was designated as martensite and the latter as bainite from the varying contrast and morphology from the optical metallography and SEM.

Figure 5(a) - 5(d) presents the inverse pole figure (IPF), image quality (IQ), phase maps, and reconstructed PAG image obtained using EBSD for the as-built HSLA samples. The IQ map is overlaid with the low-angle (5º < θ < 15º, blue lines) as well as the high-angle grain boundaries (15º < θ < 60º, red lines). Clusters of high-angle grain boundaries (red islands in the IQ maps) are observed, which signifies the presence of martensite/retained austenite (M/A) islands which are typical microstructural features of low-carbon low-alloy steels [53]. The formation of M/A islands can be attributed to the presence of microalloying elements such as Nb, Mo, and Mn that are added to compensate for the low carbon content to impart strengthening effects to the HSLA steels. These elements dissolve in the austenite and shift the austenite decomposition to lower temperatures which lead to stabilization of retained austenite. Due to carbon enrichment in the retained austenite, it partially transforms to martensite, hence, forming the M/A islands [54].

Though martensite has a body-centered tetragonal (BCT) structure, it will be considered as BCC in the analysis since the BCT structure is not available in the EBSD structure database. The IQ curve for the BCC phase (Fig. 5(e)) is found to be a single large asymmetric peak consisting of overlapping peaks. This can be attributed to the minor differences in the lattice imperfections such as dislocation densities and subgrain boundaries present in bainite and martensite [55,56]. Hence, the IQ of these phases (Fig. 5(b)) are fairly close to each other and cannot be distinguished as two different peaks from the IQ curve. According to Wu *et al*. [56], an asymmetric IQ curve confirms the presence of two different microconstituents and it can be fitted with 2 peaks (high IQ and low IQ peaks) in order to calculate the amount of each phase. Accordingly, two peaks were fitted to the IQ curve for BCC, where, the high IQ peak corresponds to the bainite and the low IQ peak corresponds to the martensite. The area under high and low IQ peaks was used to quantify the amount of bainite and martensite, respectively.

The phase map (Fig. 5(c), white: FCC and blue: BCC) shows the distribution of the FCC phase in the as-built HSLA sample. It is evident that besides the M/A constituents, the remaining FCC phase is predominantly dispersed along the grain boundaries. The FCC phase within the M/A constituents as well as along the grain boundaries corresponds to the retained austenite. The amount of bainite and martensite calculated from the IQ curve (Fig. 5(e)) as well as the amount of retained austenite obtained from the phase map are 8.4%, 84.5%, and 7.1%, respectively. From the reconstructed PAG map shown in Fig. 5(d), the PAG size was measured to be 7.5 μm, which implies that fine prior austenite grains have been achieved due to the repeated heating of the melt pools during the layer by layer deposition in the LPBF process. The hardness of the as-built HSLA sample was measured to be 384±10 $HV_{0.1}$.

### *4.2. HIP HSLA steels before and after cyclic re-austenitization*

#### *4.2.1. Cooling rate simulation for HIP*

The dilatometry curves obtained after the simulation of controlled and drop cooling using quenching dilatometry are shown in Figs. 6(a) and 6(b). The observable change in the linearity of heating and cooling curves are due to the austenite transformation from the bainitic/martensitic microstructure of the as-built HSLA samples and the martensitic transformation from the austenite, respectively. Moreover, the bainitic transformation can be observed through a minor change in the linearity of the dilatation curve before the martensite starts to form. The critical transformation temperatures such as $Ac_1$ (austenite start), $Ac_3$ (austenite finish), $M_s$ (martensite start) and $M_f$ (martensite finish) temperatures were determined from the dilatometric curves using the minimum deviation method [57]. Since the slope change due to bainitic transformation is weak, the bainite start ($B_s$) temperature was estimated by analyzing the second derivative of the dilatation variation with respect to temperature to correlate the inflection point with the start of bainitic transformation. The $B_s$ temperatures were estimated to be 667 and 625°C for the samples subjected to controlled and drop cooling, respectively. The $Ac_1$ (582°C) and $Ac_3$ (754°C) temperatures based on the CALPHAD model-prediction were used for designing experiments, although some notable deviation from experiments on AM samples are observed later. It should be further noted that the choice of $Ac_1$ is at the temperature with the phase fraction of austenite to be 1%, which is considered as an observable quantity during experiments. As shown in Figs. 6(a) and 6(c), the final length of the sample does not return to the original with a notable gap that exists between the heating and cooling curves at room temperature. This may be due to the excessive residual stresses in additively manufactured parts. It is evident from the dilatometry curves that the gap between the heating and cooling curves at room temperature is larger than the decrease in relative change in length while holding the sample at 950°C. Therefore, apart from the contraction during holding, the remaining change in relative length between the heating and cooling curves at room temperature can be attributed to the relief of residual stresses that developed during the additive manufacturing process.

The optical micrographs of the controlled and drop cooled HSLA samples are shown in Figs. 6(c) and 6(d). The microstructure consists of a mixture of bainite and martensite. It can also be observed that the microstructure after controlled cooling shown in Fig. 6(c) is coarser than that obtained after drop cooling in Fig. 6(d). The amount of martensite transformed after controlled and drop cooling was calculated to be nearly 82 vol.% and 88 vol.%, respectively, based on image analysis. This observation is expected since the $B_s$ temperature for sample subjected to controlled cooling is higher and hence, increased amount of bainite will form in comparison with the drop cooled sample. Accordingly, drop cooling is found to be suitable for performing HIP of as-built HSLA steel as it will maximize the formation of martensite during HIP.

*4.2.2. Porosity characterization*

As mentioned before, since the laser melting parameters are directly taken from the default settings from the EOS machine for SS316L, the porosity in the HIP HSLA sample was measured to be ~ 0.7% at the higher end. The optical micrographs of the as-polished X-Z plane of the as-built and HIP HSLA samples are shown in Fig. 7. The as-polished surface of the as-built HSLA sample shows the presence of pores of varying sizes and shapes such as elongated and small pores, when observed under an optical microscope (Fig. 7 (a)). On the other hand, the as-polished surface of the HIP HSLA appeared to be dense without any visible pores (Fig. 7(b)). The secondary electron images of the as-polished surfaces of the as-built and HIP HSLA steels obtained using SEM are shown in Fig. 8. There are spherical (marked with red dotted circles), irregular (marked with blue solid circles) and elongated (marked with green dashed circles) pores present in the as-built HSLA samples which would have formed due to various reasons, as discussed below. Moreover, HIP has effectively removed most of the pores (Fig.

7(b)) reaching the density as high as 99.3%, although a small trace amount of submicron pores are still visible, as shown in Fig. 7(c).

In general, porosity in AM builds can be induced either due to the powder feedstock or the AM process [58]. According to Kudzal *et al*. [59], various types of porosity encountered in alloys processed using AM can be classified into Type I, Type II and Type III porosity, depending on their morphologies. Type I pores correspond to the spherical pores that form due to the entrapment of shielding gas. These pores are spherical due to the vapor pressure of the entrapped gases, which is mostly influenced by the manufacturing process and the packing density of the powder feedstock material [60]. The gas pores remain in the as-built alloys since there are no external forces to release the gas bubbles out of the melt pools [61]. Type II pores form due to the AM process-related factors such as thermal stress-induced cracking, incomplete melting of powder particles and balling due to the inability to overcome the surface tension in the melt pool [62]. These pores are irregular and flattened in shape with sharper edges with its dimensions ranging from sub-micron to macroscopic size [58]. Type III pores are caused by the lack of fusion due to incomplete melting of the powder layer into the layer beneath and hence, leading to interlayer cracking [63]. This type of cracking forms elongated pores along the melt pool boundary with low circularity. All three types of pores mentioned above were observed from the SEM micrographs of the as-polished surface of as-built HSLA steel as shown in Figs. 8(a) and 8(b) with Type I, Type II and Type III pores marked in dotted red, solid blue and dashed green circles, respectively. The formation of these pores can be attributed to the use of factory default parameters of SS 316L for printing the HSLA-100 steel.

On the other hand, only sub-micron sized spherical pores were visible in HIP HSLA steel when viewed under SEM at high magnification as shown in Fig. 8(c). It has been reported that due to the low solubility of argon in steels, small bubbles with high internal pressure formed because of the entrapment of the protective gas persist after HIP. Hence, it is proved that the HIP has led to considerable improvement in the density of as-built HSLA steels.

*4.2.3. Microstructure and mechanical properties*

The metallographic and SEM images of the HIP HSLA steels along the build direction before and after cyclic re-austenitization are shown in Fig. 9. The microstructure comprises of bainite and martensite, similar to the as-built HSLA samples (Fig. 4). However, it can be observed that there is considerable coarsening of the microstructural features due to the slower cooling rate during HIP. Further, through a comparison between Figs. 9(b) and 9(d), it is clear that cyclic re-austenitization has led to the observable grain refinement after HIP. Figure 10 shows the IPF, IQ, phase and reconstructed PAG maps for the HIP HSLA samples before (Figs. 10(a)-10(d)) and after (Figs. 10(e)-10(h)) cyclic re-austenitization. It is observed that the orientation of the grains is random without the presence of texture from the IPF maps. M/A islands are present along with the bainite and martensite with the clustering of high angle grain boundaries as seen in the IQ maps (Figs. 10(b) and 10(f)).

The phase maps shown in Fig. 10(c) and 10(g) indicate that the retained austenite phase is sparsely distributed prior to the cyclic re-austenitization, whereas, it is densely dispersed after the re-austenitization for the HIP samples. The retained austenite other than those present within the M/A islands are distributed mostly along the grain boundaries. The amount of bainite, martensite and retained austenite obtained from the IQ curves and phase maps are 33, 64.9, and 2.1% before re-austenitization and 10.8, 84.7, and 4.5% after cyclic re-austenitization, respectively, for the HIP HSLA samples. The PAG size measured from the reconstructed PAG maps (Figs. 10(d) and 10(h)) were found to be 22.5 and 8.6 μm for HIP HSLA steel before and after cyclic re-austenitization, respectively.

It is evident from these measurements that the coarsening of prior austenite grains due to the slower cooling rate during HIP has led to the formation of coarse bainite and martensite. With the application of cyclic re-austenitization heat treatment, there is a drastic reduction in the prior austenite grain size by nearly 60% in comparison with HIP HSLA steel without re-austenitization. It has been determined experimentally that the decrease in PAG size leads to a reduction in the amount of bainite formed in a given time as the bainite transformation kinetics gets altered [64]. Hence, the reduction in the PAG size along with the faster cooling rates during quenching has led to a decrease in the amount of bainite transformed during the cyclic re-austenitization process, along with the intended refinement of bainite and martensite.

The microhardness of the as-built and HIP HSLA samples before and after cyclic re-austenitization is shown in Fig. 11. It should be noted that the hardness of the HIP HSLA sample has drastically reduced by almost 20% in comparison with the as-built HSLA steel (i.e., 384±10 $HV_{0.1}$), which can be correlated with the grain coarsening and increased amount of bainite transformed as observed from the microstructure characterization. On the other hand, the microhardness of the HIP HSLA sample after cyclic re-austenitization is found to be on par with the microhardness of the as-built HSLA steel since the microstructural features were refined. This proves that a post-heat treatment involving two cycles of re-austenitization after HIP is necessary to improve the properties of the additive manufactured HSLA-100 steels.

### *4.3. Tempering of HIP HSLA steels before and after cyclic re-austenitization*

Figure 11 presents the microhardness as a function of time for the HIP samples before and after cyclic re-austenitization after tempering at 550°C. The optimum tempering time with peak hardness is 5 hours in both cases. In copper-bearing HSLA steels, Cu initially precipitates with spherical shape and metastable BCC structure from the supersaturated α-Fe solid solution due to the similar size of Fe and Cu atoms [15]. At the peak-aged condition, the coherent BCC clusters attain an average diameter of 1 – 5 nm with nearly 50 at.% of Cu remaining in the solid solution [16]. When a critical size (4 – 12 nm) is reached, there is a reduction in strain and interfacial energy that induces a secondary transformation to an intermediate close-packed structure called 9R leading to a decrease in the strengthening effects [65,66].

According to Gagliano *et al*. [15], overaged Cu precipitates are observed at the peak aged condition, and the decrease in strengthening associated with the formation of these overaged precipitates are compensated by the secondary hardening from the NbC that forms during austenitization. However, $M_2C$ precipitates were not observed in the peak aged condition, possibly due to the absence of Mo in the HSLA steels considered in their work. Subsequently, Mulholland *et al*. [17] studied the precipitation after tempering HSLA steels using atom probe tomography and reported that the precipitation of Cu resulted in the clustering of carbon atoms which eventually leads to the heterogeneous nucleation of $M_2C$ particles within the Cu precipitates. Based on this observation, it was postulated that the presence of co-located $M_2C$ particles within Cu is found to impede the coarsening of Cu precipitates since Cu is insoluble in $M_2C$ carbide [17]. Hence, the peak hardness achieved during tempering of AM HSLA-100 steels is likely to be achieved because of the co-precipitation of strengthening particles such as Cu and $M_2C$ with optimum size, shape, and coherency, which eventually leads to precipitation strengthening.

Figure 12 shows the optical and SEM micrographs for the tempered HIP HSLA steels before and after cyclic re-austenitization with peak hardness (5 hours). The microstructure of both the alloys consist of tempered martensite and bainite, which can be differentiated by their phase contrast and morphology. The IPF, IQ and phase maps obtained from EBSD for the tempered HIP HSLA steels before and after cyclic re-austenitization with peak hardness are shown in Fig. 13. According to the IPF maps presented in Fig. 13(a) and 13(d), it can be found that there is no preferred orientation. The clustering of high angle grain boundaries that signifies the presence of M/A islands can be observed from the IQ plots as shown in Figs. 13(b) and 13(e). From the phase maps (Figs. 13(c) and 13(f)), it is evident that the FCC phase is present in the M/A islands as well as the grain boundaries. The amount of bainite, martensite and retained austenite calculated from the EBSD measurements are 31%, 66.3% and 2.5% for tempered HIP HSLA steel, and 10.2% 85% and 4.8% for tempered HIP HSLA steel after cyclic re-austenitization, respectively. As shown in Fig. 11, the hardness of tempered HIP HSLA steel after cyclic re-austenitization is higher than that of the tempered HIP HSLA steel before cyclic re-austenitization, for different tempering time. The presence of the lesser amount of bainite in the tempered HIP HSLA steel after cyclic re-austenitization has led to the increase in the hardness by ~10% in the peak aged condition, in comparison with the tempered HIP HSLA steel before cyclic re-austenitization.

The stress-strain curves and the tensile properties of the as-built and HIP samples before and after cyclic re-austenitization tempered for 5 hours to achieve peak aging condition are shown in Figs. 14(a) and 14(b), respectively. The yield strength of the as-built HSLA steel was found to be 106 ksi (735 MPa), which is greater than the targeted yield strength of HSLA-100 steels. However, the elongation was found to be only 13%, which is below the design target as 16% (see Fig. 1). Due to the preferentially oriented grains along the direction of highest thermal gradient, the microstructure of the components made by AM are anisotropic (see Fig. 5), resulting in anisotropic mechanical properties [67]. The anisotropy in the mechanical properties of as-built parts has been reported for several materials [67–70]. Hence, it is expected that the as-built HSLA steel would also exhibit varying mechanical properties in different orientations. Moreover, it has been explained using slip theory analysis that melt pool boundaries have significant influence on the microscopic slipping, macroscopic plastic behavior and fracture mode, and are responsible for the anisotropy and poor ductility in parts fabricated using LPBF [68]. Therefore, the low elongation observed in the as-built HSLA steels can be attributed to the slipping of the melt pool boundaries perpendicular to the loading direction.

There is a considerable decrease in the yield strength and ultimate tensile strength as well as an increase in the elongation in the tempered HIP HSLA steel before (or without) cyclic re-austenitization. This behavior can be attributed to the high fraction of bainite formed and the coarse microstructure after HIP. The highest yield strength of ~117 ksi (812 MPa) was achieved for the tempered HIP HSLA steel after cyclic re-austenitization. Moreover, there is a considerable improvement in the elongation for the tempered HIP HSLA steel after cyclic re-austenitization, in comparison with the as-built HSLA steel. This is because of the isotropic microstructure achieved after HIP and cyclic re-austenitization processes, as it can be observed from Figs. 10 and 13, compared to the anisotropic microstructure of the as-built HSLA steel (Fig. 5). It can be envisaged that the tensile properties will also be isotropic for HIP HSLA steels before and after cyclic re-austenitization due to the isotropic microstructure. The ultimate tensile strengths of the as-built sample and HIP HSLA steel after cyclic re-austenitization are similar. Hence, it is evident that the tensile properties of the tempered HIP HSLA steel after cyclic re-austenitization is well above the design targets that are specified for yield strength and elongation in the systems design chart for AM HSLA-100 steel as shown in Fig. 1. The presence of a refined microstructure along with possible precipitation strengthening during tempering has led to the drastic improvement in the strength of the HIP HSLA steel after cyclic re-austenitization. This necessitates the need for a distortion-free post-heat treatment after HIP to achieve dense additive manufactured HSLA steels with improved mechanical properties.

## 5. Summary and conclusions

- In the present work, an effective post-heat treatment process was designed for additively manufactured copper-bearing HSLA steel. HSLA-100 steel was fabricated using the LPBF method successfully for the first time. Microstructure characterization of as-built HSLA steel revealed the presence of bainite, martensite, retained austenite and M/A islands.

- Since the factory default building parameters for 316L stainless steel were adopted for fabricating the HSLA steel, the as-built HSLA sample consisted of high porosity which required HIP for densification. Through quenching dilatometry, drop cooling was found to be more suitable for HIP as it can maximize the martensite formation. The porosity reduced from 3% to 0.7% after HIP, along with considerable grain coarsening and a higher fraction of transformed bainite.

- Cyclic re-austenitization was performed after HIP to achieve grain refinement which necessitated the need for post-heat treatment after HIP to attain improved mechanical properties. Further strengthening was introduced by tempering the HIP HSLA steel before and after re-austenitization. A significant increase in the yield strength and ductility of the tempered HIP HSLA steel after cyclic re-austenitization was achieved in comparison with the as-built HSLA steel.

- With the application of the designed post-heat treatment process, the design targets were achieved for AM HSLA-100 steels. More time-dependent properties and corrosion resistance will be studied after further optimization of the laser melting parameters for more condensed builds. Despite this, such work demonstrates the requirement for post-heat treatment design that relies on microstructure engineering in order to achieve improved mechanical properties for AM components.

**CRediT Author statement**

**Soumya Sridar:** Validation, Formal analysis, Investigation, Data Curation, Writing - Original Draft; **Yunhao Zhao:** Validation, Investigation, Writing - Review & Editing; **Kun Li:** Validation, Investigation, Writing - Review & Editing; **Xin Wang:** Investigation, Writing - Review & Editing; **Wei Xiong:** Conceptualization, Methodology, Investigation, Writing - Review & Editing, Supervision, Project administration, Funding acquisition.

**Acknowledgment**

The authors would like to gratefully acknowledge the financial support received from the Office of Naval Research (ONR) Additive Manufacturing Alloys for Naval Environments (AMANE) program (Contract No.: N00014-17-1-2586).

**Figure captions:**

Figure 1.   Systems design chart for HSLA-100 steels fabricated using additive manufacturing.

Figure 2.   Calculated equilibrium phase fraction as a function of temperature using the TCFE9 database of Thermo-Calc for the pre-alloyed HSLA steel powder. The $Ac_1$ temperature is the temperature at which the austenite phase fraction is greater than 1 %.

Figure 3.   Temperature profile for the HIP processes with (a) controlled cooling process (b) drop cooling followed by cyclic re-austenitization with two cycles.

Figure 4.   (a) Optical and (b) SEM micrographs of as-built HSLA steel. (B: Bainite, M: Martensite)

Figure 5.   (a) IPF, (b) IQ (c) Phase, and (d) Reconstructed PAG boundary maps as well as (e) IQ curve for BCC obtained from EBSD for as-built HSLA steel.

Figure 6.   Dilatometry curves showing the calculated transition temperatures and their corresponding optical micrographs for (a, c) controlled and (b, d) drop cooling applied using quenching dilatometry to as-built HSLA samples (B: Bainite; M: Martensite).

Figure 7.   As-polished surfaces of (a) as-built and (b) HIP HSLA steels viewed using optical microscopy.

Figure 8.   Secondary electron images of the as-polished surfaces of (a, b) as-built and (c) HIP HSLA steels. Type I, Type II and Type III porosities observed in the as-built HSLA sample are marked in dotted red, solid blue and dashed green circles, respectively.

Figure 9.   Microstructure of HIP HSLA samples viewed under both optical microscope and SEM  (a, b) before, and (c, d) after cyclic re-austenitization (B: Bainite, M: Martensite).

Figure 10.  Microstructure images under EBSD. (a) IPF, (b) IQ, (c) phase, and (d) reconstructed PAG boundary maps for HIP HSLA steels before cyclic re-austenitization. (e) IPF, (f) IQ, (g) phase, and (h) reconstructed PAG boundary maps for HIP HSLA steels after cyclic re-austenitization.

Figure 11.  Microhardness variation vs. tempering time for as-built HSLA steel as well as HIP HSLA samples before and after cyclic re-austenitization.

Figure 12.  Optical and SEM micrographs of HIP HSLA steels tempered for 5 hours (a, b) before and (c, d) after cyclic re-austenitization (TM: Tempered Martensite; B: Bainite).

Figure 13.  IPF, IQ, and phase maps of HIP HSLA steels tempered for 5 hours (a-c) before and (d-f) after cyclic re-austenitization obtained from EBSD.

Figure 14.  (a) Stress-strain curves and (b) Tensile properties for as-built HSLA sample and HIP HSLA  steels tempered for 5 hours before and after cyclic re-austenitization.

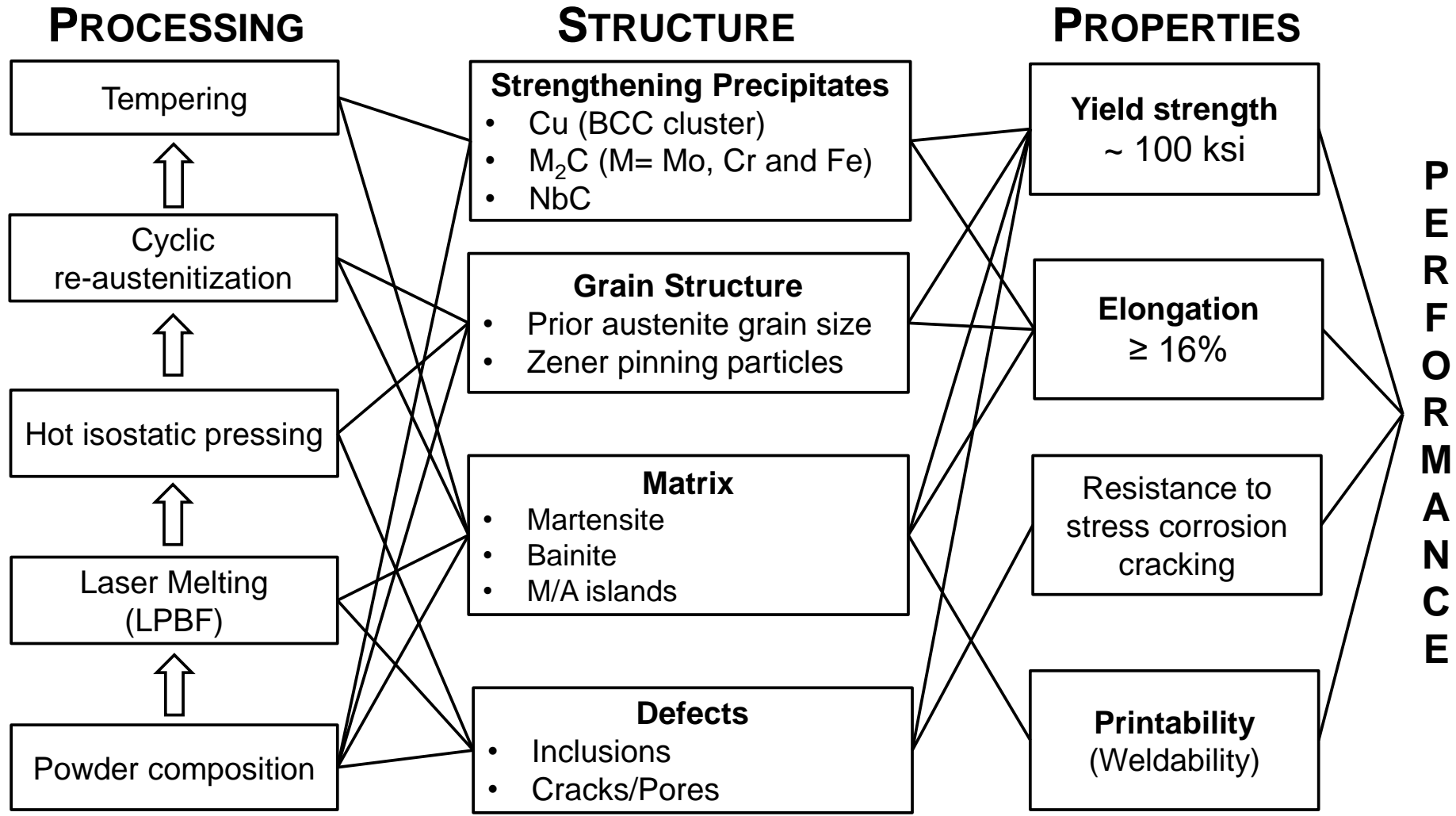

Figure 1. Systems design chart for HSLA-100 steels fabricated using additive manufacturing.

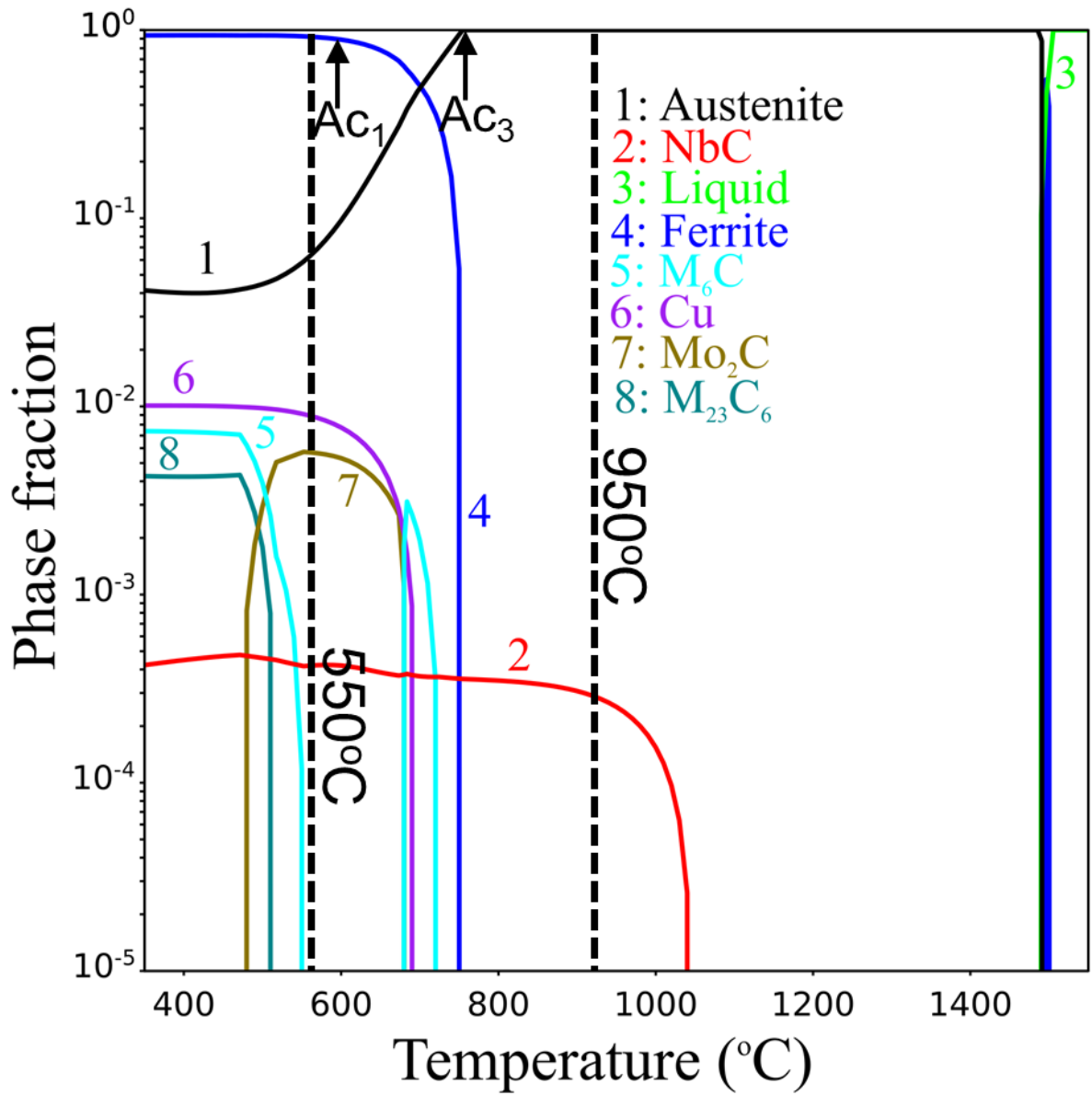

Figure 2. Calculated equilibrium phase fraction as a function of temperature using the TCFE9 database of Thermo-Calc for the pre-alloyed HSLA steel powder. The $Ac_1$ temperature is estimated with the austenite phase fraction greater than 1 %.

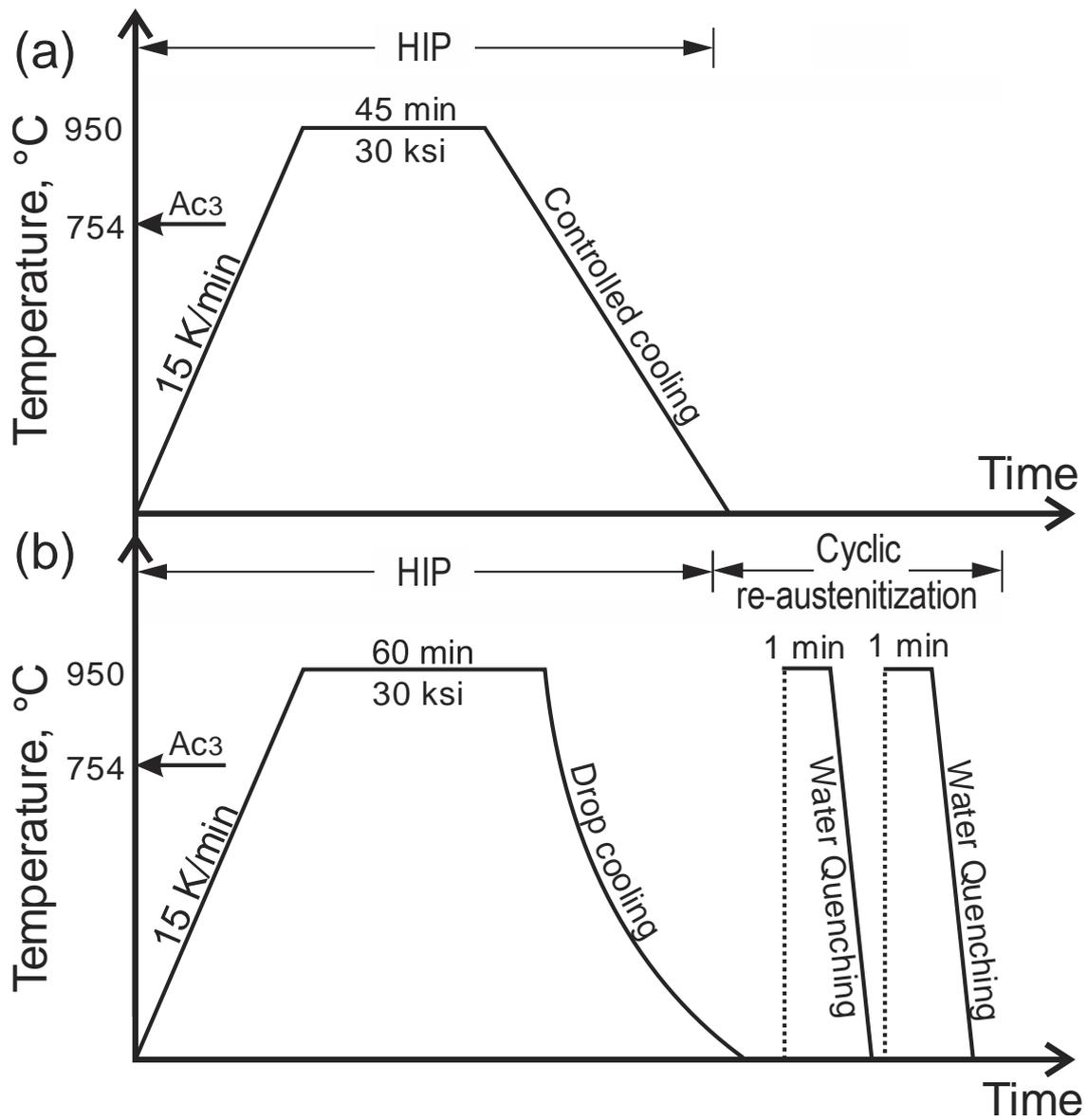

Figure 3. (a) Temperature profile for the HIP processes with (a) controlled cooling process with a rate of 10 K/min, and (b) drop cooling followed by cyclic re-austenitization with two cycles. During drop cooling of HIP, the estimated average cooling rate from 800 to 500°C is 15 K/min.

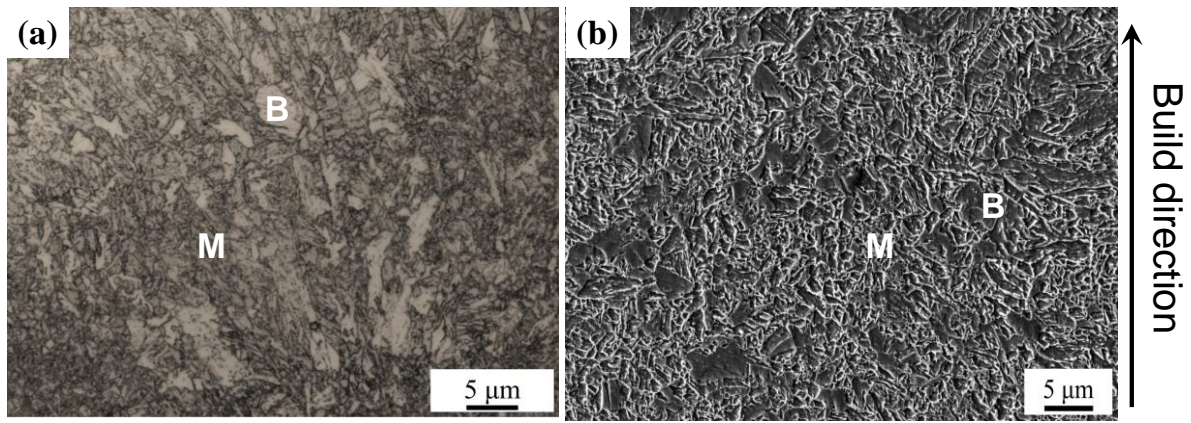

Figure 4. (a) Optical and (b) SEM micrographs of as-built HSLA steel. (B: Bainite, M: Martensite)

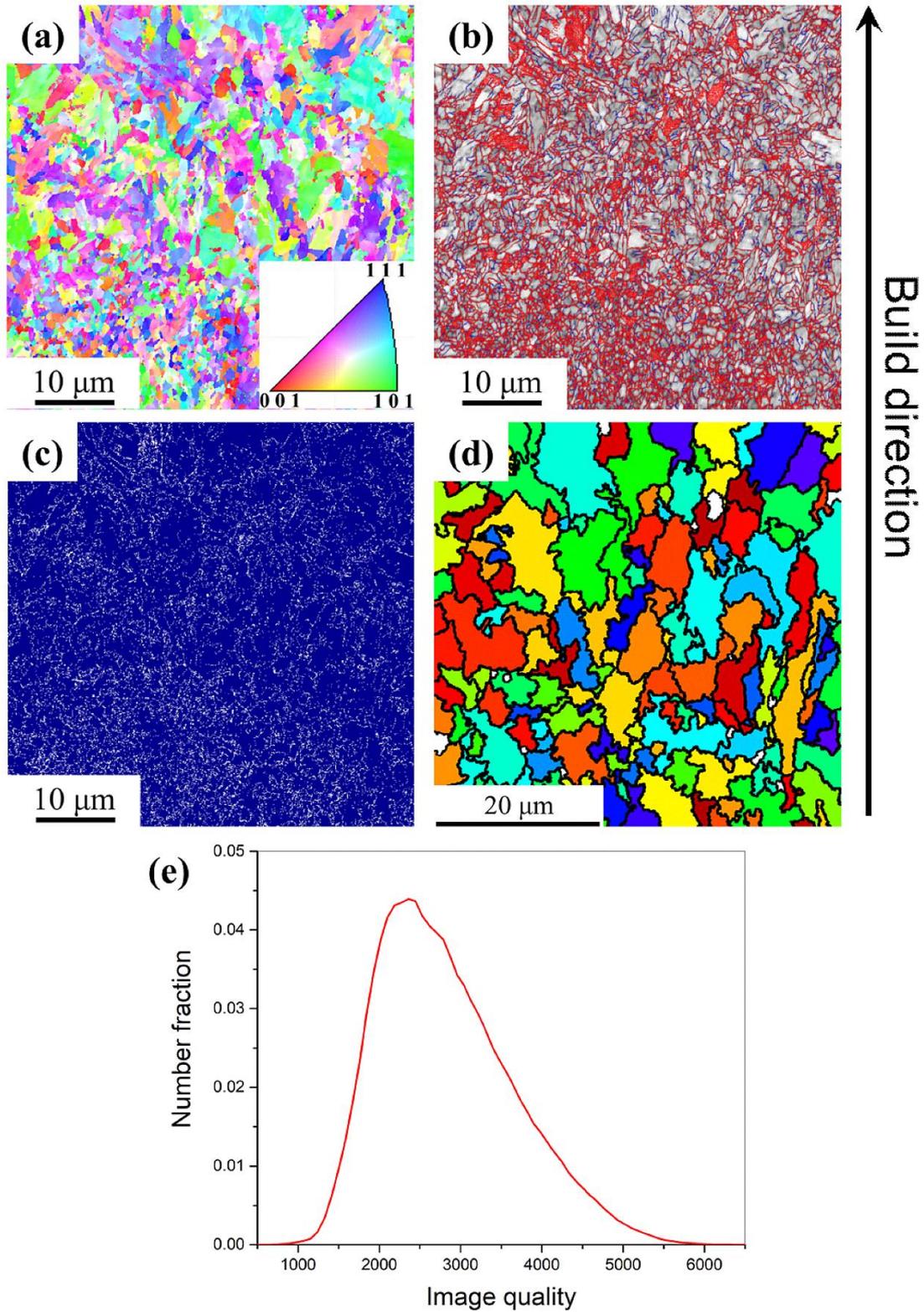

Figure 5. (a) IPF, (b) IQ (c) Phase, and (d) Reconstructed PAG boundary maps as well as (e) IQ curve for BCC obtained from EBSD for as-built HSLA steels.

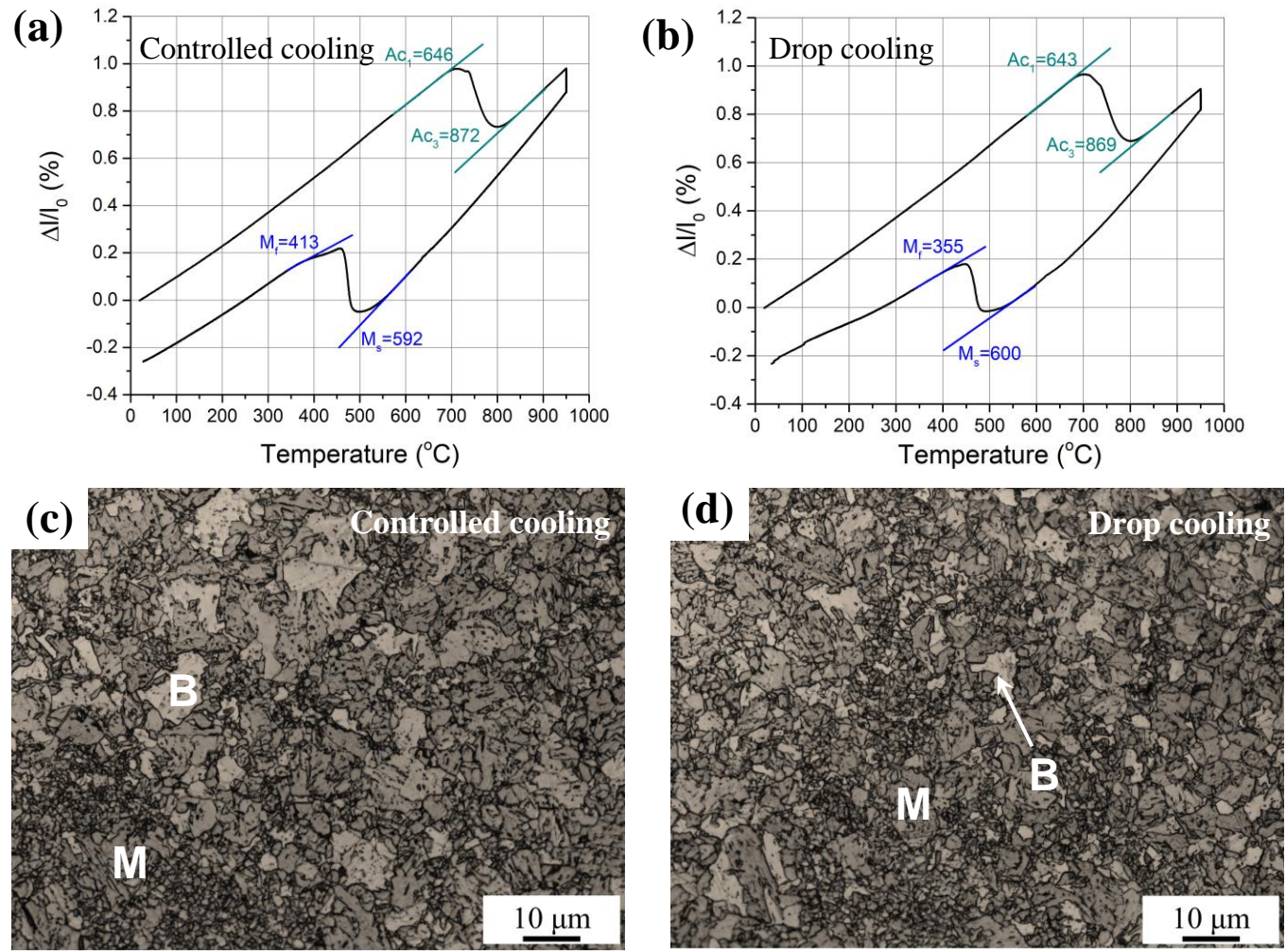

Figure 6. Dilatometry curves showing the calculated transition temperatures and their corresponding optical micrographs for (a, c) controlled and (b, d) drop cooling applied using quenching dilatometry to as-built HSLA samples (B: Bainite; M: Martensite).

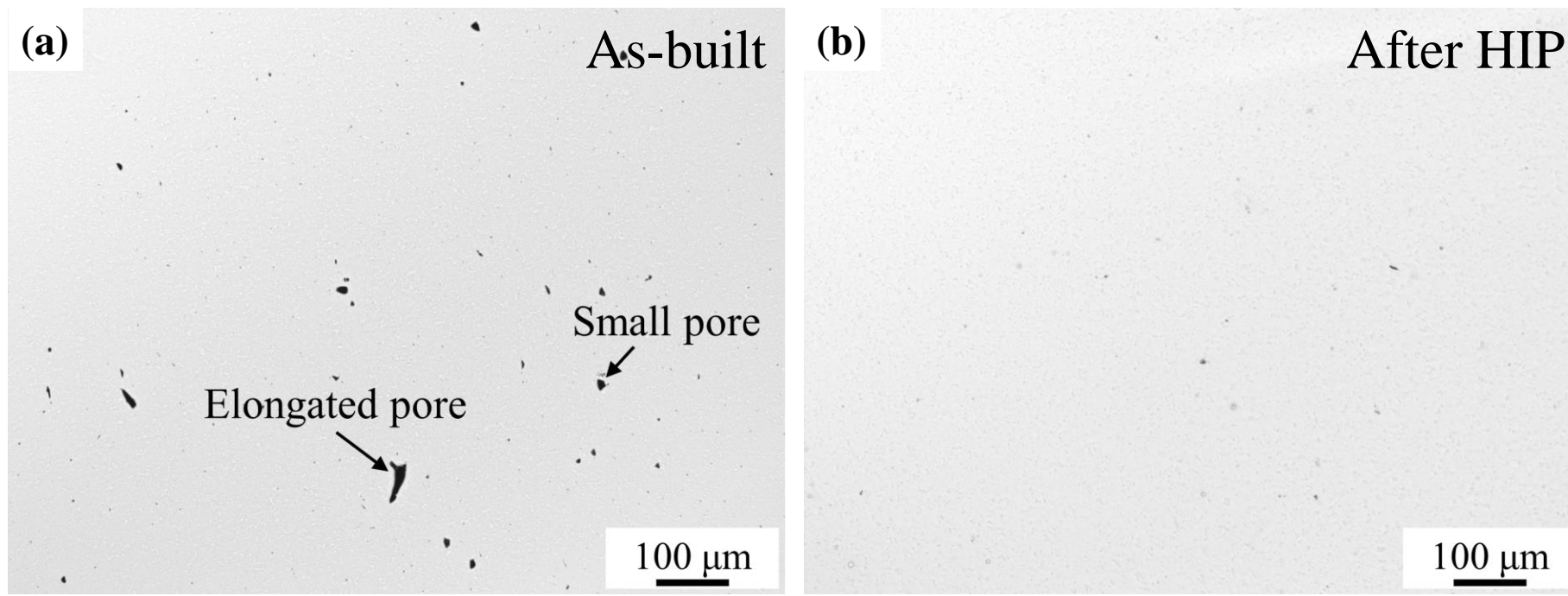

Figure 7. As-polished surfaces of (a) as-built and (b) HIP HSLA steels viewed using optical microscopy.

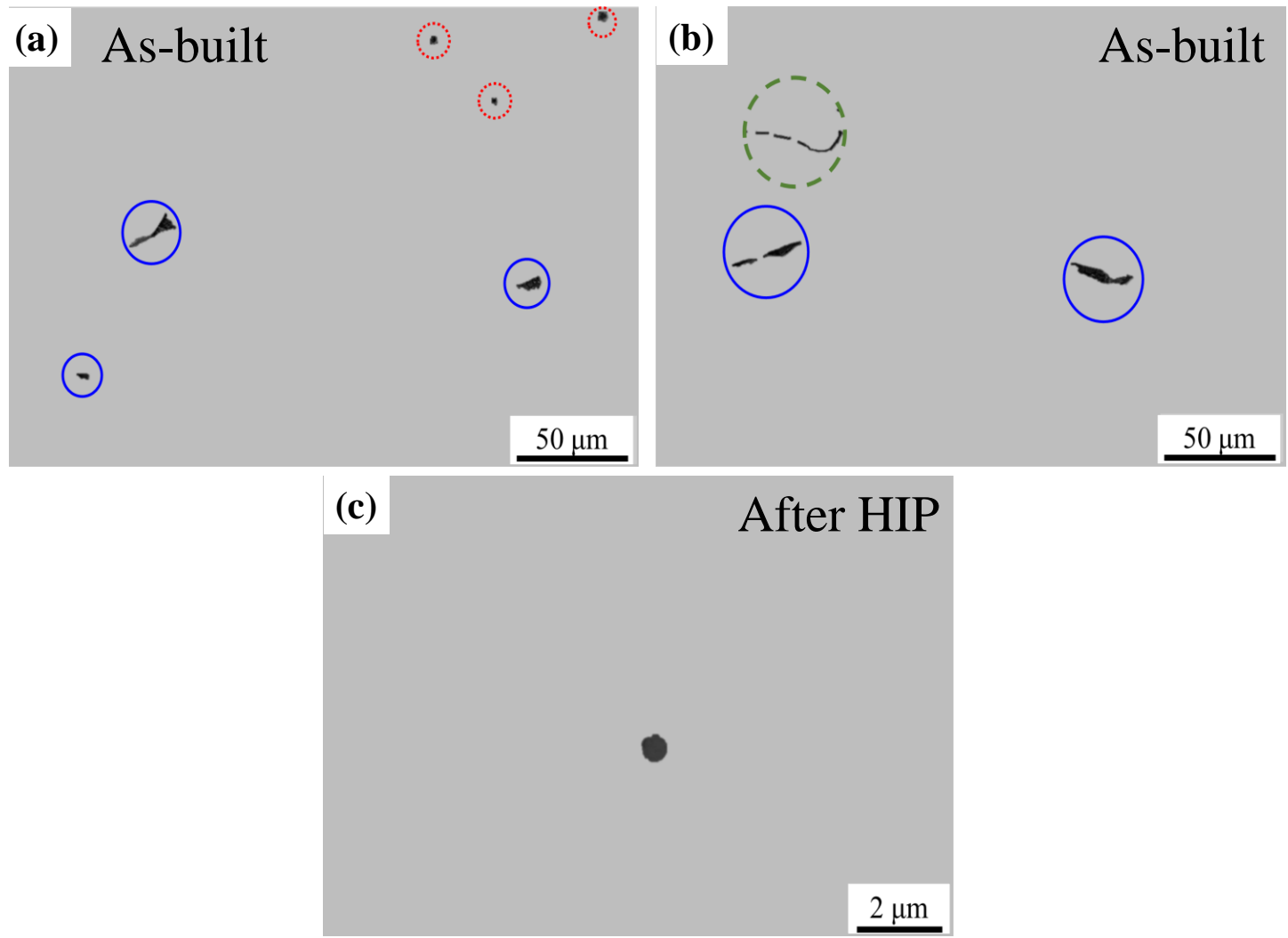

Figure 8. Secondary electron images of the as-polished surfaces of (a, b) as-built and (c) HIP HSLA steels. Type I, Type II and Type III porosities observed in the as-built HSLA sample are marked in dotted red, solid blue and dashed green circles, respectively.

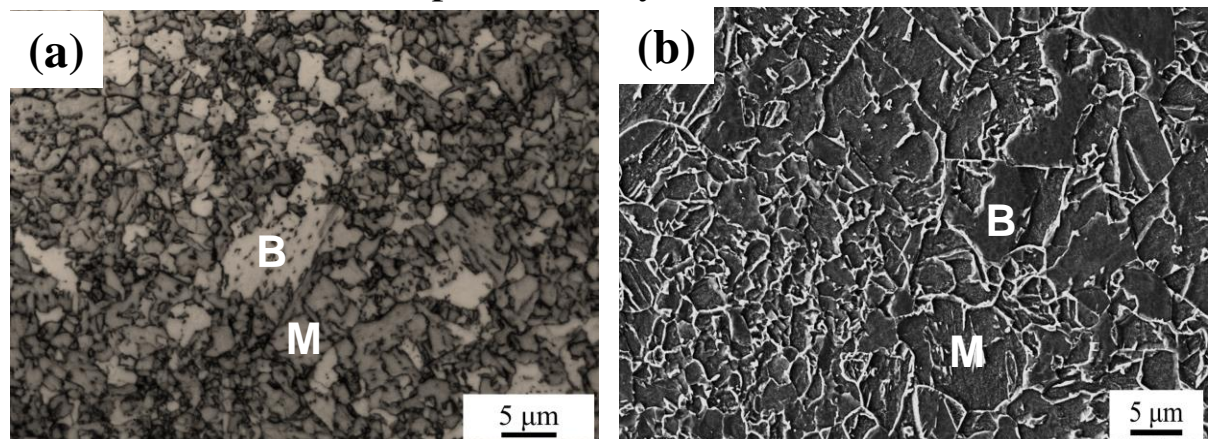

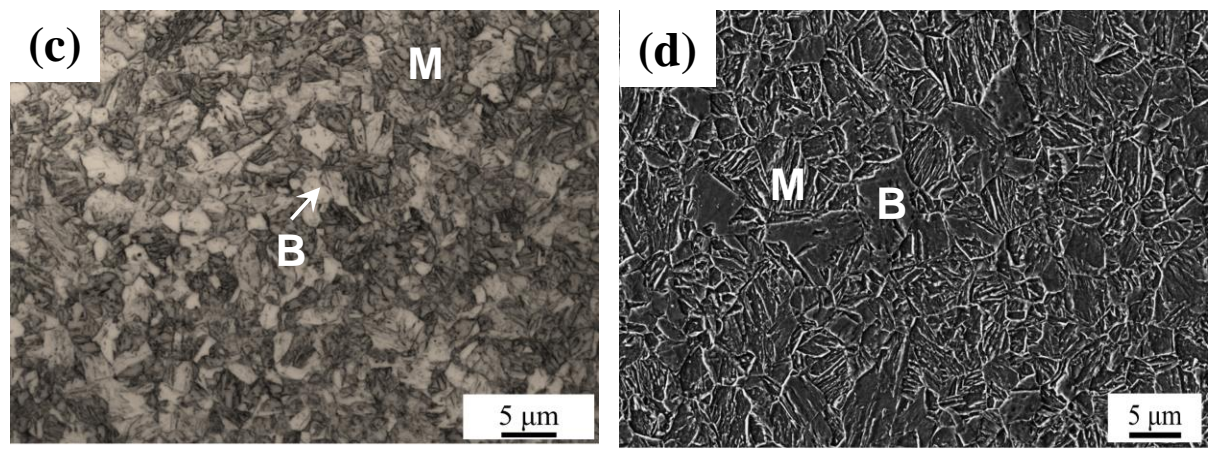

Figure 9. Microstructure of HIP HSLA samples viewed under both optical microscope and SEM (a, b) before, and (c, d) after cyclic re-austenitization (B: Bainite, M: Martensite).

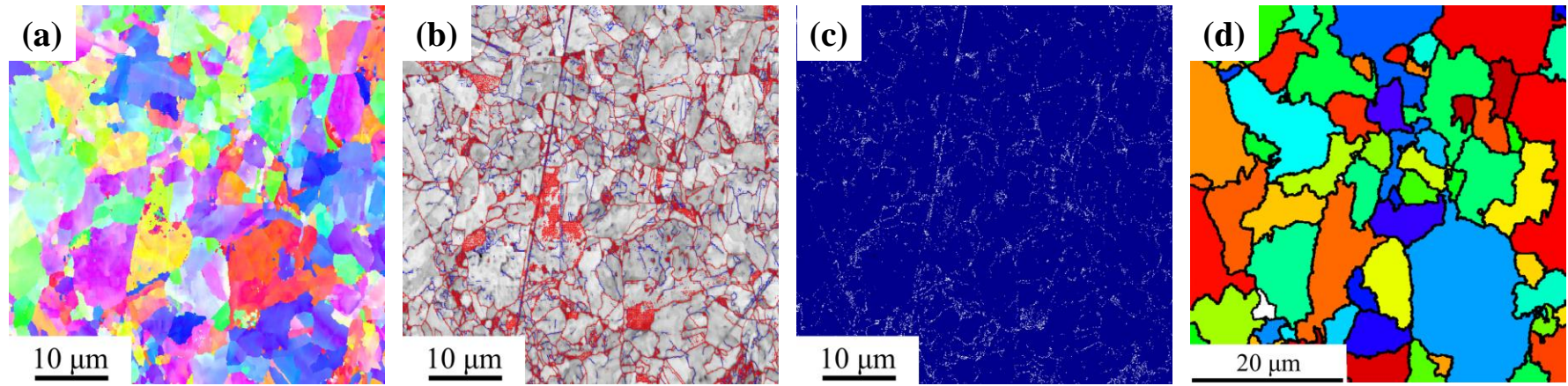

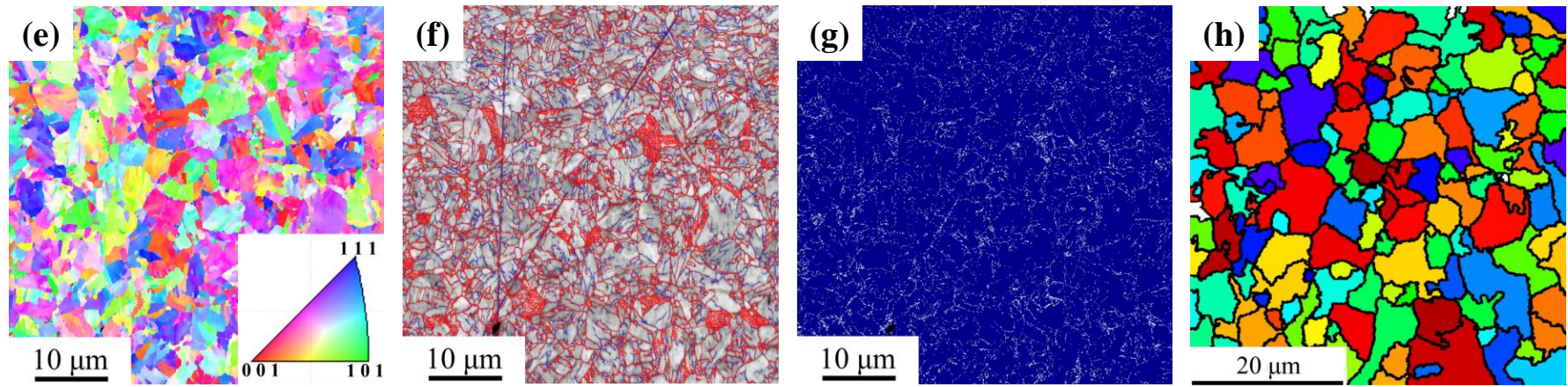

Figure 10. Microstructure images under EBSD. (a) IPF, (b) IQ, (c) phase, and (d) reconstructed PAG boundary maps for HIP HSLA steels before cyclic re-austenitization. (e) IPF, (f) IQ, (g) phase, and (h) reconstructed PAG boundary maps for HIP HSLA steels after cyclic re-austenitization.

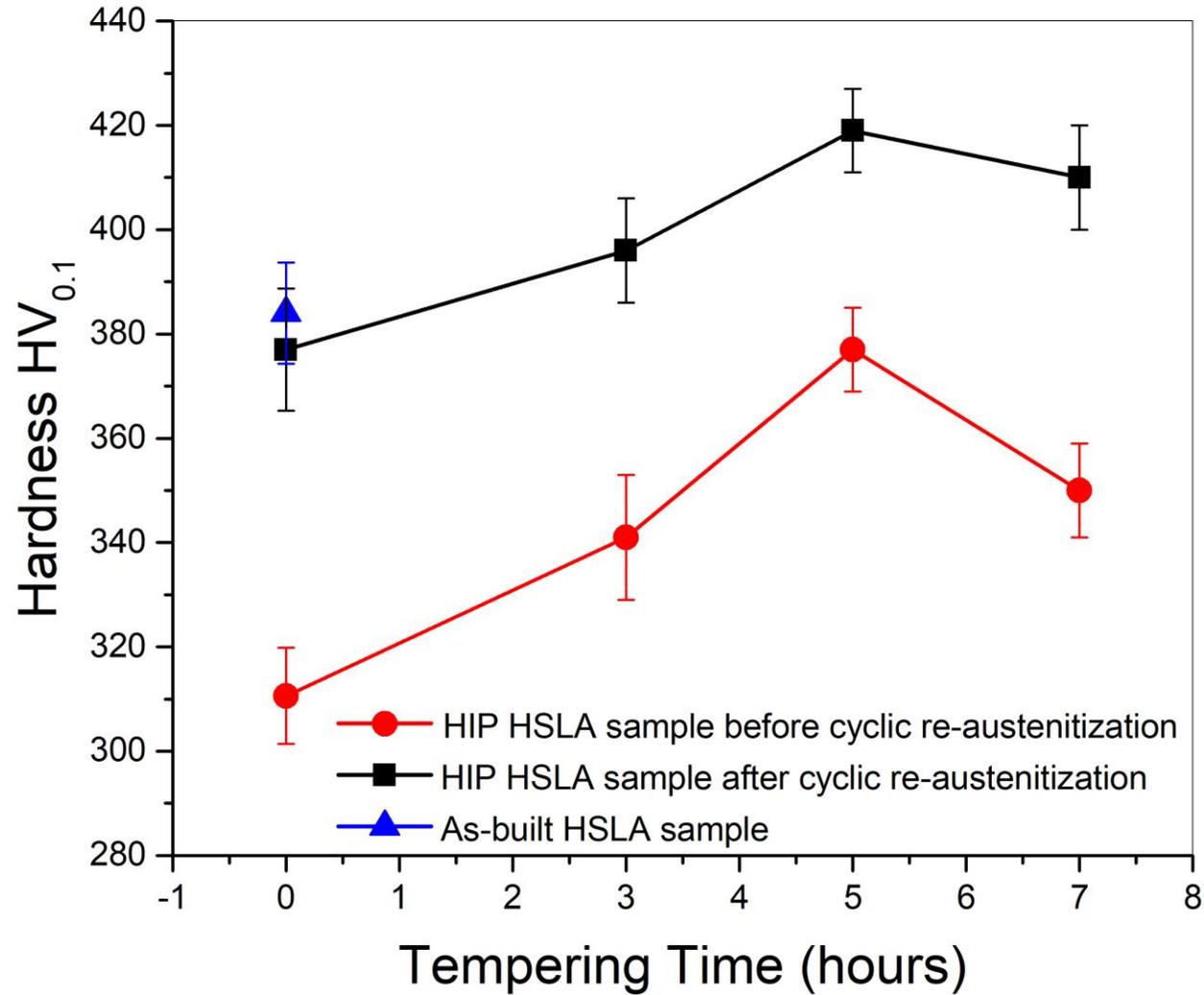

Figure 11. Microhardness variation vs. tempering time for as-built HSLA steel as well as HIP HSLA samples before and after cyclic re-austenitization.

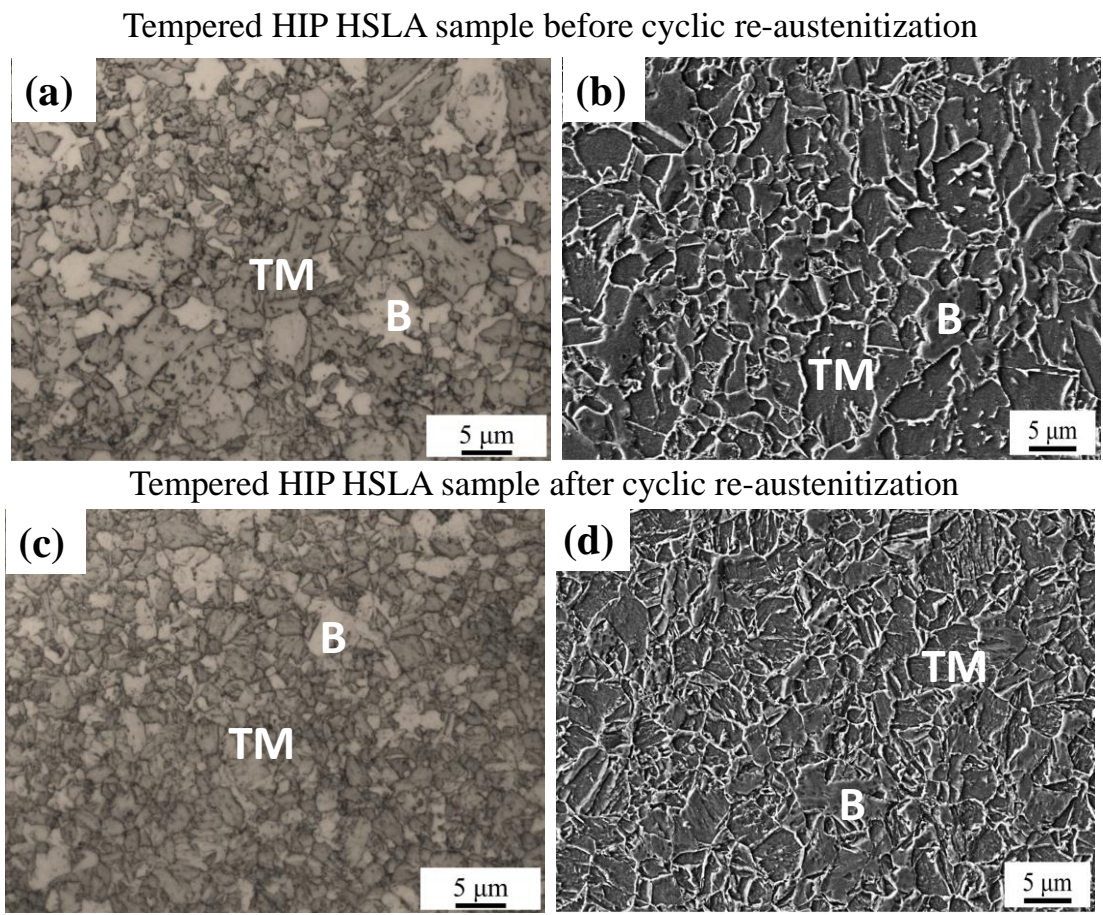

Figure 12. Optical and SEM micrographs of HIP HSLA steels tempered for 5 hours (a, b) before and (c, d) after cyclic re-austenitization (TM: Tempered Martensite; B: Bainite).

Tempered HIP HSLA sample before cyclic re-austenitization

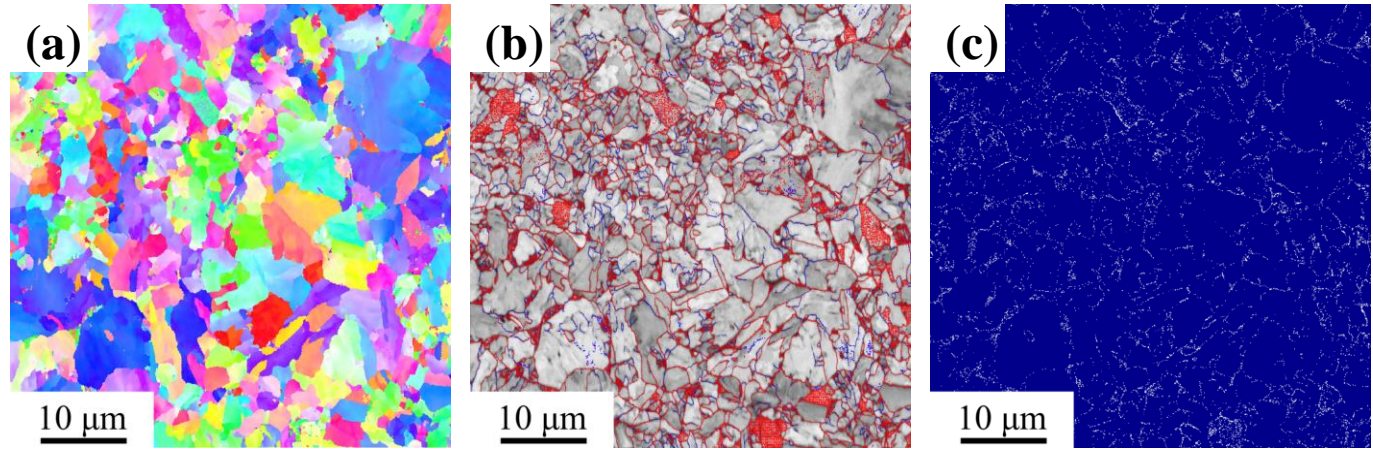

Tempered HIP HSLA sample after cyclic re-austenitization

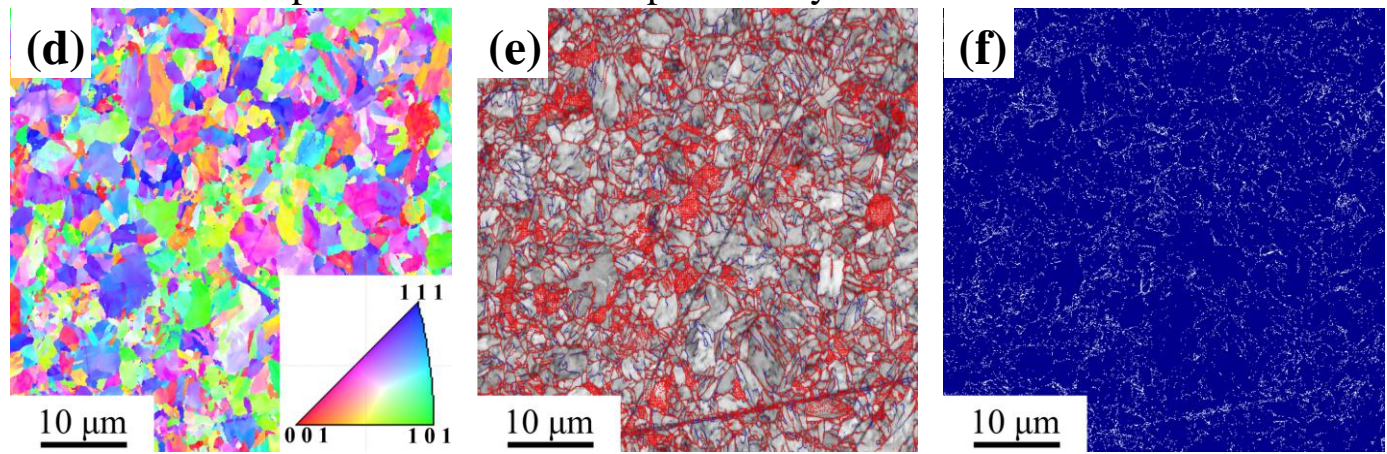

Figure 13. IPF, IQ and phase maps of HIP HSLA steels tempered for 5 hours (a-c) before and (d-f) after cyclic re-austenitization obtained from EBSD.

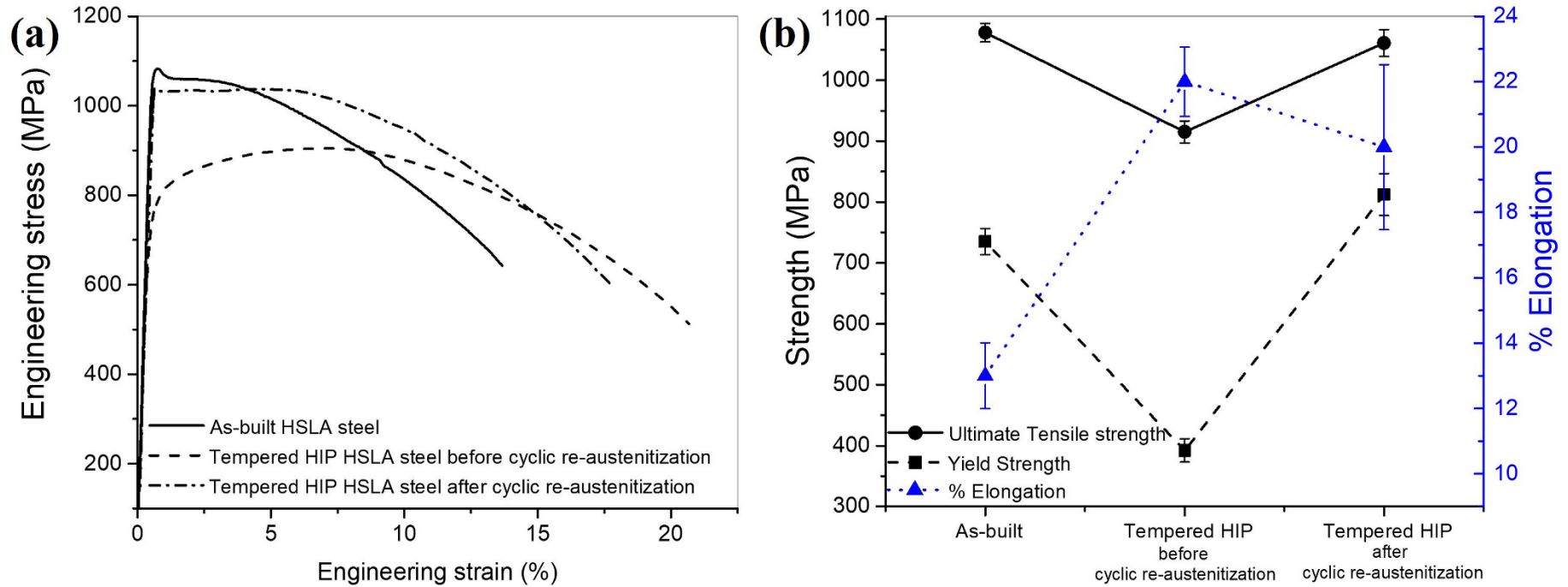

Figure 14. (a) Stress-strain curves and (b) Tensile properties for as-built HSLA sample and HIP HSLA steels tempered for 5 hours before and after cyclic re-austenitization.